# Observation of surface layering in a nonmetallic liquid

Haiding Mo,[1] Guennadi Evmenenko,[1] Sumit Kewalramani,[1] Kyungil Kim,[1]

Steven N. Ehrlich[2] and Pulak Dutta[1]

[1]Department of Physics & Astronomy, Northwestern University, Evanston, IL 60208 USA

[2]National Synchrotron Light Source, Brookhaven National Laboratory, Upton, NY 11973 USA

68.10.-m, 68.15.+e, 68.35.Ct

Oscillatory density profiles (layers) have previously been observed at the free surfaces of liquid metals, but not in other isotropic liquids. We have used x-ray reflectivity to study a molecular liquid, tetrakis(2-ethylhexoxy)silane. When cooled to $T/T_c \approx 0.25$ (well above the freezing point for this liquid), density oscillations appear at the surface. Lateral order within the layers is liquid-like. Our results confirm theoretical predictions that a surface-layered state will appear even in dielectric liquids at sufficiently low temperatures, if not preempted by freezing.

Until recently, the structure of a liquid surface was thought to be well understood:[1] the density changes monotonically from the density of the liquid to that of the vapor. However, Rice et al.[2] predicted in 1973 that there would be density oscillations (layers) at liquid metal surfaces. This theory assigned an essential role to the electron gas in causing an abrupt transition between the conducting liquid and nonconducting vapor, resulting in an effective wall potential. Starting in 1995, X-ray scattering experiments have established that such smectic-like order does exist near the surfaces of a number of liquid metals and metallic alloys.[3-7] (This phenomenon is distinct from surface freezing,[8] which is a pretransition effect seen in certain liquids only just above the bulk freezing point, and where the surface layer is frozen, i.e. laterally ordered.) There have also been many X-ray studies of the surface profiles of classical dielectric liquids, including alkanes,[9] ethanol,[10] toluene,[11] polymers and polymer solutions[11,12] and water,[13,14] but these have shown no evidence of surface layers away from the freezing point.

The presence of an electron gas is the obvious feature that distinguishes metals from dielectrics, but in practice there are other differences. The surface tension of most liquid metals is very high (e.g. 500mN/m for Hg), which means that the surface is very smooth. Dielectric liquids have much lower surface tensions (<100mN/m, and frequently <40mN/m) and therefore much rougher surfaces. Dielectric liquids are known to form layers at the surfaces of smooth hard substrates.[15,16] Therefore, one possibility is that dielectric liquids would show surface layering if only their free surfaces were not so rough.[17] Another possibility was raised by Chacón et al,[18] who concluded from simulations that surface layering will appear below about $0.2T_c$ where $T_c$ is the critical temperature. In many liquid metals, the critical temperature is high and the melting temperature low; thus low values of $T/T_c$ can be reached within the liquid phase. For example, layering is seen in mercury at room temperature, where $T \approx 0.15 T_c$. However, most classical dielectric liquids freeze at much higher $T/T_c$ (e.g. water freezes at $0.42 T_c$). In general, a low freezing point and high critical point are consequences of a broad and shallow pair potential; Li and Rice[19] have labeled such liquids "Madrid liquids".

The question, therefore, is whether the observed difference between metallic and dielectric liquid surfaces is due to the electron gas, or whether surface layering is a universal property of liquids that may in some cases be preempted by freezing or masked by surface roughness.

We have studied the surface of a molecular liquid, tetrakis(2-ethoxyhexoxy)silane (TEHOS), using X-ray reflectivity. TEHOS is an isotropic (non-liquid-crystalline) dielectric liquid. The molecule consists of one Si and four O atoms in the center, surrounded by four saturated branched alkanes forming a "wax coating" that makes the molecules nonreactive and roughly spherical. It is used in outdoor devices such as transformers and solar cells because it does not freeze in winter and has very low evaporation loss. Neither the freezing point nor the critical point has been precisely determined, but viscosity measurements[20] have shown that it is a fluid down to at least 219K. Using X-ray scattering in transmission, and differential scanning calorimetry, we have found no evidence of a bulk phase transition down to 190K. The boiling point is 467K at 1mm Hg.[20] Using the Clausius-Clapeyron equation to estimate the normal boiling point, and then using the normal boiling point to estimate the critical temperature,[21] we find that $T_c$ is ~950K. Thus TEHOS is like a liquid metal in that it has a low melting point and a high critical point.

In order to easily cool the liquid, we prepared ~5000Å films of TEHOS supported on silicon substrates. These were formed by putting a few drops of liquid on the substrate, allowing the liquid to spread, and then draining the excess.[16] These samples can then be mounted in a closed-cycle refrigerator and then oriented as needed in a diffractometer. A somewhat similar method has been used previously to study liquid helium.[22] The liquid film thickness is much larger than relevant length scales (surface roughness, molecular dimensions, etc.) We saw no measurable changes in film thickness over at least 12 hours, which confirms that the evaporation rate is very low.

It is known that TEHOS forms layers near smooth silicon surfaces.[15,16] To avoid seeing features in the reflectivity due to interface layering, which is not of interest in the present study, we prepared and used substrates with RMS surface roughness >20Å. Yu *et al*.[16] have found, and we have verified, that the scattering features due to interfacial layers can no longer be seen when the substrate surface is rough. The internal interface contributes only a diffuse background to the reflectivity data. To prepare the rough silicon wafers, we started with polished silicon wafers and cleaned them using the procedure as described in Ref. 16. We then etched them for 3 min. in 6% hydrogen fluoride solution. This cleaning and etching procedure was repeated once and then the etched silicon wafers were cleaned once again.

TEHOS was purchased from GLEST Inc with a purity of >95% and used as supplied. Specular x-ray reflectivity studies were performed at MATRIX (Beam Line X18A, National Synchrotron Light Source) and MUCAT (Sector 6, Advanced Photon Source) using a conventional four-circle diffractometer. The beam size was ~0.8mm vertically and ~1mm horizontally. The momentum resolution was ~0.006Å$^{-1}$. The samples were mounted on the cold head of a closed-cycle refrigerator and covered with a beryllium radiation shield, which helps to keep the sample temperature uniform. The cold head and the sample were then sealed under vacuum with a beryllium can. The whole system was pumped with a molecular turbo pump to maintain a vacuum. Before collecting data, the sample was kept at the desired temperature for at least 30 min. for the system to reach equilibrium. In addition to specular scans, slightly off-specular 'background' scans were performed and subtracted from the specular data, thus removing the scattering from all diffuse sources including that from the rough liquid-solid interface.

Figure 1 shows the specular reflectivity R divided by the Fresnel reflectivity $R_F$ at several temperatures. At 237K (and at higher temperatures, not shown here), the scans are featureless. At lower temperatures, distinct reflectivity oscillations are seen, indicating that there is some structure in the interfacial electron density $\rho(z)$ averaged over the surface plane. The change in the reflectivity data appears at the same temperature whether we are going up or down, and does not have any detectable dependence on age or temperature history of the sample, X-ray exposure, etc. The temperature threshold corresponds to $T/T_c \approx 0.25$. This value of $T/T_c$ is lower than has been reached during other any other classical dielectric liquid studies.

Both the Patterson functions and model-independent fits[23] (not shown) confirm that the reflectivity changes at 227K and below result from the formation of surface layers. The fits we show here use the 'semi-infinite series of Gaussians' model frequently used to fit reflectivity data from liquid metal surfaces.[5,14,24] However, a series of equally spaced layers will not fit our data. Guided by model-independent fits, we have modified the model function so that there is a different spacing ($d_0$) and a different relative density ($r$) for the first two layers:

$$\frac{\rho}{\rho_{bulk}} = r\sum_{n=0}^{1}\frac{d_0}{\sqrt{2\pi}\sigma_n}e^{-\frac{(z+nd_0)^2}{2\sigma_n^2}} + \sum_{n=2}^{\infty}\frac{d_1}{\sqrt{2\pi}\sigma_n}e^{-\frac{[z+(n-2)d_1+2d_0]^2}{2\sigma_n^2}}$$

where $\sigma_n^2 = \sigma_o^2 + n\bar{\sigma}^2$, so that the width of the layer increases with layer number (i.e. with depth below the surface). Further, the starting width $\sigma_0$ is a combination of two terms: $\sigma_0^2 = \sigma_i^2 + \sigma_{cw}^2$, where $\sigma_{cw}$ is due to thermal capillary waves and $\sigma_i$ is an intrinsic term due to all other factors. Note that $\sigma_{cw}$ is not a variable parameter: we have measured the surface tension as a function of temperature using the Wilhelmy-plate method in the range 265-305K, extrapolated it to the temperatures of interest,[25] and used this to calculate $\sigma_{cw}$.[26] Most of the fitting parameters show no noticeable temperature dependence: $d_0$=8.4±0.2Å, $d_1$=10.3±0.1Å, $\bar{\sigma}$=2.0±0.1Å and r=1.15±0.02. However, $\sigma_0$ increases with temperature, as expected, from 3.8 Å at 197K to 4.2Å at 227K.

The solid lines in Fig. 1 show that good fits are obtained with this model. (For the featureless reflectivity curves at higher temperatures, we used the familiar one-layer model with an error-function-broadened interface). Fig. 2 shows the fitted densities. The dashed lines are the full best-fit density functions; the solid lines are the same functions except with $\sigma_{cw}$ = 0. In other words, the solid lines show what the surface profiles would look like if they had not been broadened by thermal capillary waves.

Could impurity segregation to the surface be responsible for the onset of layering? Since impurities are mobile, even the highest-purity samples of any metallic or dielectric liquid have far more impurities than are needed to cover the surface if they were to migrate there. However, impurities will go to the surface only if this reduces the surface energy, which will result in an increase in surface width. In Fig. 3 we show the top surface width as a function of temperature. In order to fairly compare the function used at 227K and below to that used at 237K and above, we have defined the width as the distance from $\rho/\rho_{bulk}= 0.9$ to $\rho/\rho_{bulk}= 0.1$; thus the absolute values will differ from widths calculated using other definitions. It can be seen in Fig. 3 that the surface width is monotonic; there is no increase in surface width when layering appears at lower temperatures. Also, since impurities are unlikely to be the same size as TEHOS molecules, the lateral short-range order in an impurity layer should be different from that of bulk TEHOS. Fig. 4 shows grazing-incidence (surface sensitive) X-ray scattering data at 212K, compared with bulk diffraction data (in transmission) at the same temperature. At the incident angle of 0.05º (the critical angle for liquid TEHOS is ~0.10º), the X-rays penetrate only ~100Å into the liquid, which means that scattering from the ~30Å layered region is a significant fraction of the observed scattering. The bulk and surface data are not expected to be look identical because of the very different scattering geometries, but there are no sharp features in either curve (i.e. there is only liquid-like order, just as at liquid metal surfaces), and the broad features have about the same positions and widths in both cases (i.e. there is no evidence of a different molecular species).

Shpyrko *et al* have investigated the role of surface width by studying liquid potassium[5] and water.[14] The surface tension of liquid potassium is much lower than that of most liquid metals and comparable to that of water (~70 mN/m). From an analysis of differences in the X-ray reflectivity and diffuse scattering data, they conclude that there is surface layering in liquid potassium but none in water. The surface tension of TEHOS is

~32 mN/m at room temperature, and ~35.5 mN/m (extrapolated) at 227K; both these values are lower than those for water and potassium in the studies referred to above. Thus the observation of surface layering is not correlated with whether the surface tension is high or low, but neither is it correlated with whether the liquid is a metal or a dielectric. On the other hand, potassium and water were not (and cannot be) studied at the same $T/T_c$: in the potassium study, $T/T_c \approx 0.15$, whereas for the water study, $T/T_c \approx 0.45$. Recall that the onset of layering in TEHOS is at $T/T_c \approx 0.25$. Thus a low $T/T_c$ is necessary, although there is as yet no evidence of a universal $T/T_c$ threshold.

In summary, we have observed surface layering in a non-liquid-crystalline, nonmetallic liquid. The presence of an electron gas is not necessary.

This work was supported by the US National Science Foundation under Grant No. DMR-0305494. Use of the National Synchrotron Light Source (NSLS) is supported by the U.S. Department of Energy under Contract No. DE-AC02-98CH10886. The MATRIX beam line at NSLS is supported by the US Department of Energy. The Advanced Photon Source (APS) is supported by the U.S. Department of Energy under Contract No. W-31-109-Eng-38. The MUCAT sector at the APS is supported by the U.S. Department of Energy through the Ames Laboratory under Contract No. W-7405-Eng-82.

Figure Captions:

Fig1: Normalized specular reflectivity for ~5000Å TEHOS films on rough silicon wafers at different temperatures. At 227K and below, the reflectivity data develop structure indicating a change in the surface density profile. Lines are best fits using the electron density profiles shown in Fig. 2.

Fig.2 The dashed lines show best-fit electron densities as functions of distance from the surface, for ~5000Å TEHOS films on rough silicon wafers at different temperatures. The solid lines show the density profiles with capillary broadening removed (see text).

Fig 3 Liquid surface width vs. temperature. The surface width is defined in a fitting-function-independent way as the distance from $\rho/\rho_{bulk}= 0.9$ to $\rho/\rho_{bulk}= 0.1$ in each of the dashed lines in Fig. 2; thus the absolute values will differ from widths calculated using other definitions.

Fig. 4 Grazing incidence (in-plane) scattering from a TEHOS surface (o), compared to scattering in transmission from a bulk TEHOS sample (●) . The two curves show the same features, indicating that there is only liquid-like lateral order at the surface.

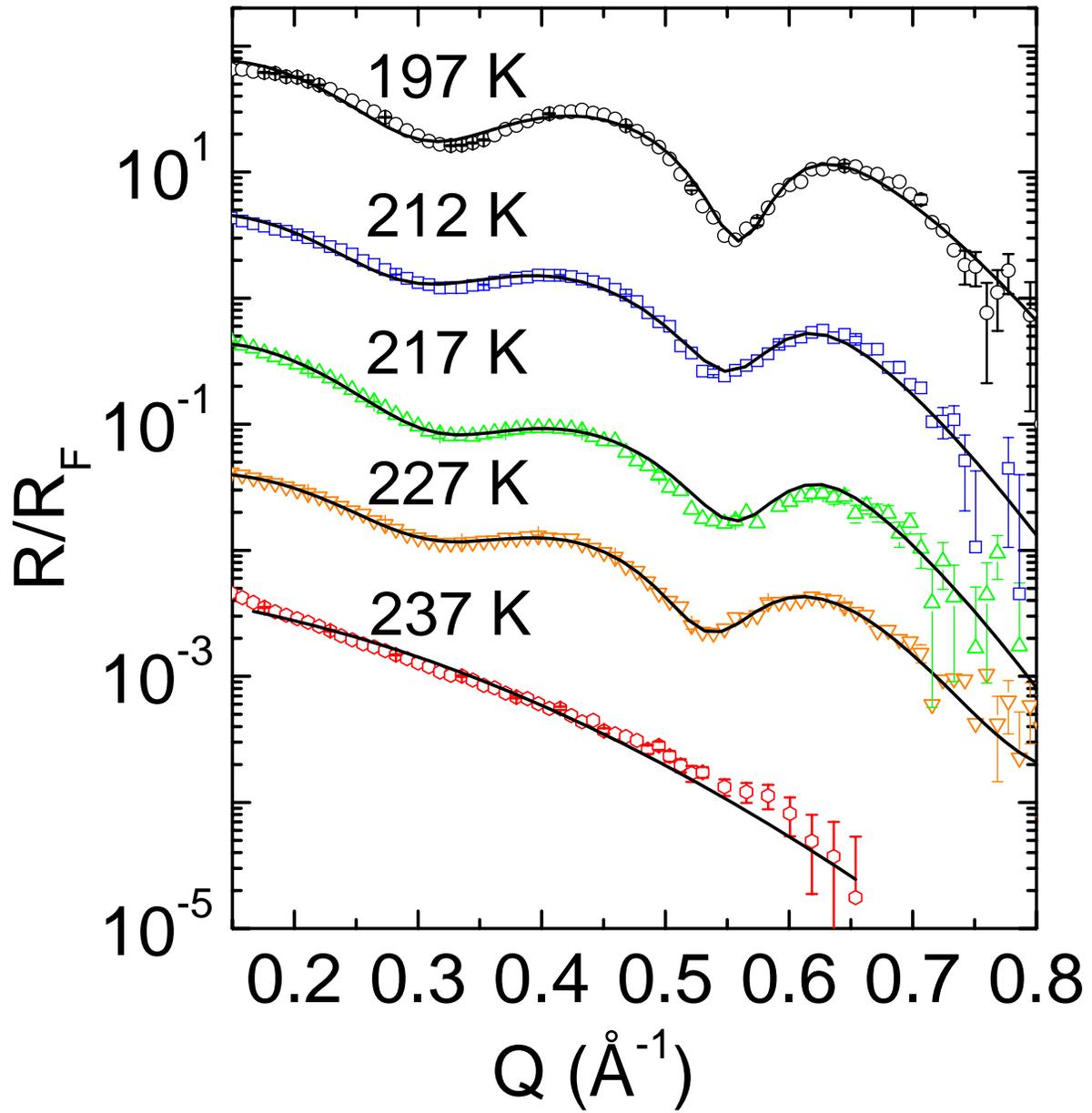

Fig. 2

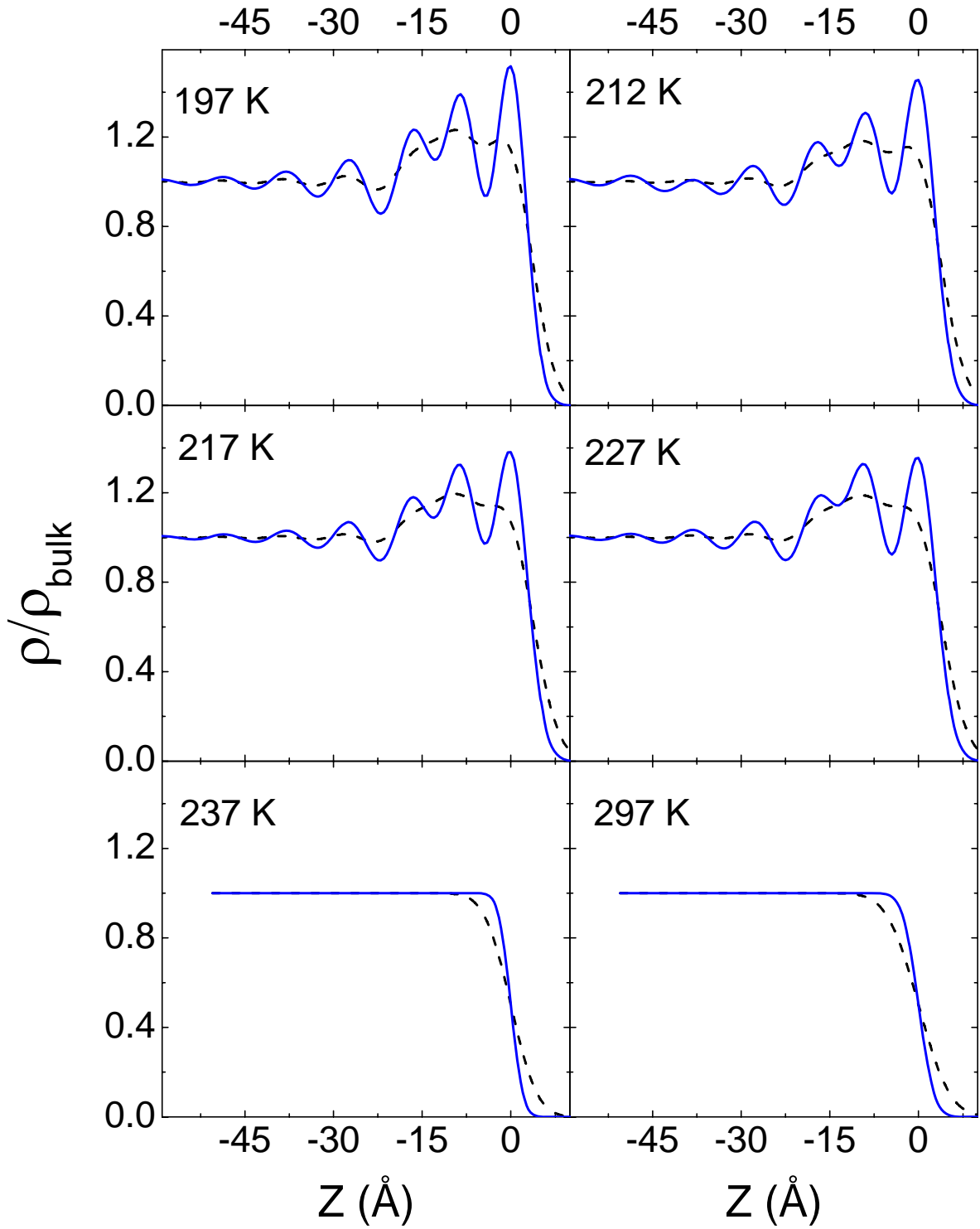

Fig. 3

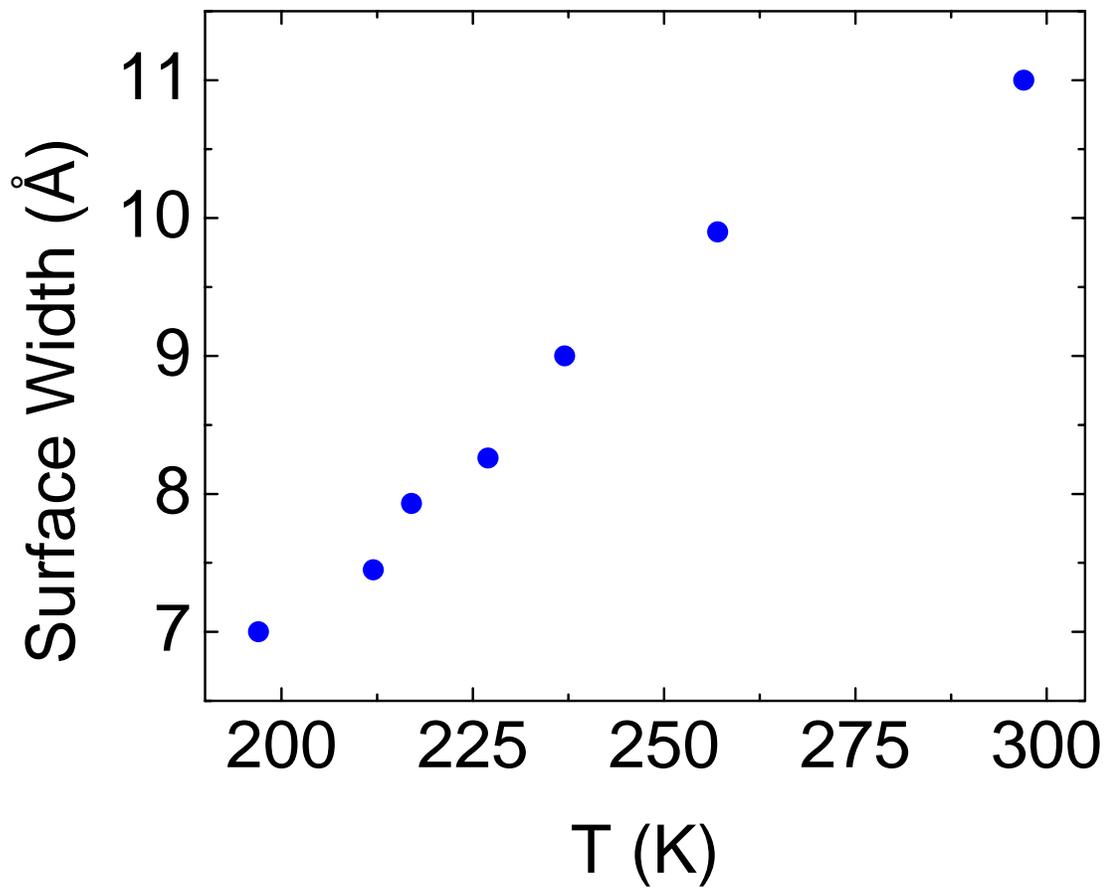

Fig. 4

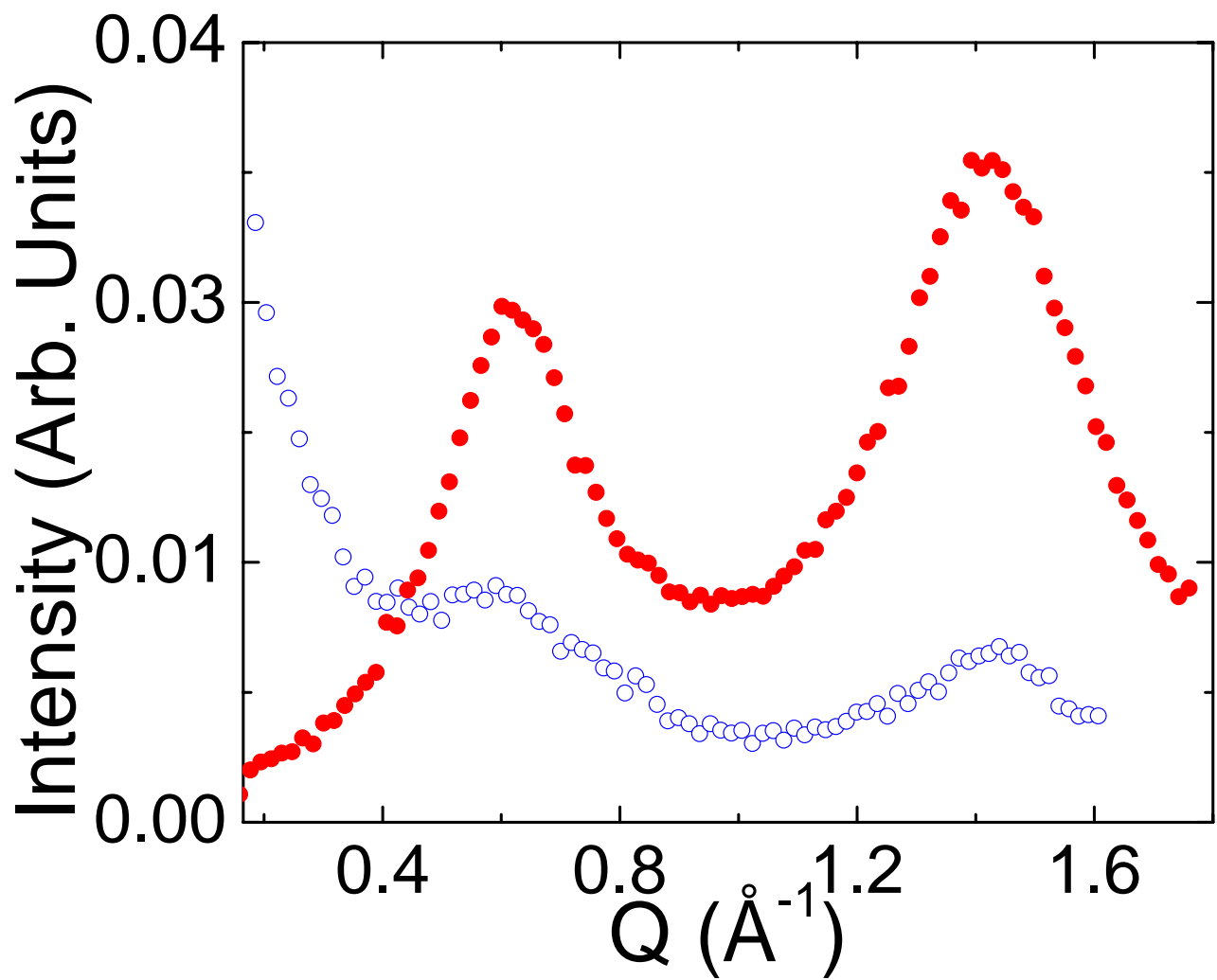